\documentstyle[fleqn,twoside,gc]{article}

\heads{S. Ogushi}
      {HOLOGRAPHIC ENTROPY ON THE BRANE}

\begin{document}
\twocolumn[
\Arthead{6}{2000}{4 (24)}{1}{10}

\Title{HOLOGRAPHIC ENTROPY ON THE BRANE \yy
       FROM DS/CFT CORRESPONDENCE}


   \Author{Sachiko Ogushi\foom 1 } 
          {Yukawa Institute for Theoretical Physics, 
          Kyoto University, Kyoto 606-8502, JAPAN}              
          

\Abstract
{We discuss the relationship between the entropy of de Sitter (dS) 
Schwarzschild space and that of the CFT which
lives on the brane by using Friedmann-Robertson-Walker 
(FRW) equations and Cardy-Verlinde formula.  
The cosmological constant appears on the brane 
with time-like metric in dS Schwarzschild background.  
On the other hand, in case of the brane with 
space-like metric in dS Schwarzschild background, 
the cosmological constant of the brane
does not appear because we can choose brane tension 
to cancel it.  We show that when the time-like brane crosses 
the  black hole horizon of dS Schwarzschild black hole, the entropy of 
the CFT exactly agrees with the black hole entropy 
of 5-dimensional dS background.  This report is
based on the work \cite{SOG}.}



]  
\email 1 {ogushi@yukawa.kyoto-u.ac.jp}

\def\be{\begin{equation}}
\def\ee{\end{equation}}
\def\bea{\begin{eqnarray}}
\def\eea{\end{eqnarray}}
\def\nn{\nonumber \\}
\def\e{{\rm e}}

The holographic duality which connects $d+1$--dimensional 
gravity in Anti-de Sitter (AdS) background with 
$d$-dimensional conformal field theory (CFT) 
has been discussed vigorously for some years.  
The one of the evidences for the existence of the AdS/CFT 
correspondence is that the isometry of $d+1$-dimensional AdS space 
$SO(d,2)$ is identical with the conformal symmetry of 
$d$-dimensional Minkowski space.  
Recently much attention has been paid 
for the duality between de Sitter (dS) gravity and CFT 
by the analogy of the AdS/CFT correspondence, which
is called dS/CFT correspondence, because  
the isometry of $d+1$-dimensional de Sitter space, $SO(d+1,1)$, 
exactly agrees with the conformal symmetry of 
$d$-dimensional Euclidean space.    
Thus it might be natural to expect the correspondence 
between $d+1$-dimensional gravity in de Sitter space and 
$d$-dimensional Euclidean CFT.  
The dS/CFT correspondence is initiated by C.M. Hull \cite{HL},
A. Strominger \cite{AS} and E. Witten \cite{EW}.  
Following their works, much attention has been paid for dS/CFT 
correspondence \cite{SOG},\cite{SO2},\cite{CI}.

Moreover the holographic principle 
between the radiation dominated Friedmann-Robertson-Walker (FRW) 
universe in $d$-dimensions and same dimensional CFT with a dual
$d+1$-dimensional AdS description was studied by E. Verlinde \cite{EV}.  
Especially, we can see the correspondence between
black hole entropy and the entropy of the CFT
which is derived by making the appropriate
identifications for FRW equation with
the generalized Cardy formula.
The Cardy formula is originally the entropy 
formula of the CFT only for 2-dimensions \cite{CD},
while the generalized Cardy formula expresses 
that of the CFT for any dimensions \cite{EV}.  
From the point of brane-world physics \cite{RS},
the CFT/FRW relation sheds further light on the study
of the brane CFT in the background of AdS Schwarzschild 
black hole \cite{SV}.  There was much activity on the studies
of related questions \cite{SO2},\cite{SO1},\cite{SD},\cite{DF},\cite{DF1}.  

The purpose of this report is the further study of the CFT
in de Sitter (dS) Schwarzschild background 
guided by the analogy of AdS Schwarzschild background.  
The investigation of dS brane in dS Schwarzschild 
background in terms of FRW equations has been initiated in 
the work \cite{SD}.  The important difference between 
AdS space and dS space is the sign of cosmological constant.  
In case of the brane with time-like (Minkowski) metric 
on AdS Schwarzschild background \cite{SV},   
the cosmological constant does not appear because
we can choose brane tension to cancel it.  But 
it is impossible for dS Schwarzschild background 
with the positive cosmological constant.  
We will see that the cosmological constant always appears in 
FRW equations deduced from time-like brane trajectory in
dS Schwarzschild background
\footnote{The systematic method deriving brane cosmological 
equation for a brane embedded in a bulk with a cosmological 
constant has been first examined by the works \cite{DF}.}.  
It is interesting to note that the brane with space-like (Euclidean) metric 
in dS Schwarzschild background, the cosmological constant 
of the brane does not appear for the same reason in case of AdS 
Schwarzschild background.
From the point of view of the dS/CFT correspondence, 
the investigation of space-like brane is more 
interesting than that of time-like brane.  

Furthermore we argue the entropy of the brane CFT which
is derived by using generalized Cardy formula for both 
time-like and space-like branes.  
We will see that when the time-like brane crosses 
the black hole horizon of dS Schwarzschild black hole, 
the CFT is identical with the black hole entropy 
of 5-dimensional dS background as it happens in the
AdS/CFT correspondence.

\section{FRW equations in the background of de Sitter
Schwarzschild black hole}

We first consider a 4-dimensional time-like 
brane in 5-dimensional dS Schwarzschild background.  
From the analogy of the AdS/CFT correspondence, 
we can regard that 4-dimensional CFT exists on the brane 
which is the boundary of the 5-dimensional 
dS Schwarzschild background.  
The dynamics of the brane is described by the boundary action:
\bea
{\cal L}_{b} ={-1 \over 8\pi G_{5}}\int_{\partial {\cal M}}
\sqrt{-g}{\cal K}+{\kappa \over 8\pi G_{5}}
\int_{\partial {\cal M}} \sqrt{-g}\ .
\eea
Here $G_5$ is 5-dimensional bulk Newton constant, 
$\partial {\cal M}$ denotes the surface of the brane, $g$ is the 
determinant of the induced metric on $\partial {\cal M}$, 
${\cal K}_{ij}$ is the extrinsic curvature, and ${\cal K}={\cal K}^{i}_{i}$, 
$\kappa$ is a parameter related to tension of the brane.  

From this Lagrangian, we can get the equation 
of motion of the brane as \cite{SV}:
\bea
\label{eom}
{\cal K}_{ij}={\kappa \over 2}g_{ij}\ ,
\eea
which implies that $\partial {\cal M}$ is a brane of 
constant extrinsic curvature.  
The bulk action is given by 5-dimensional Einstein
action with cosmological constant.
The dS Schwarzschild space is one of the exact solutions
of bulk equations of motion and can be written in the
following form,
\bea
\label{SAdS}
ds^{2}_{5}&=&\hat G_{\mu\nu}dx^\mu dx^\nu \nn
&=& -\e^{2\rho} dt^2 + \e^{-2\rho} dr^2 
+ r^2 d \Omega_{3}^2 \ ,\nn
\e^{2\rho}&=&{1 \over r^{2} }\left( -\mu + r^{2}  
- {r^4 \over l^2} \right) \ .
\eea
Here $l$ is the curvature radius of dS and $\mu$ is the
black hole mass.  In case of AdS Schwarzschild 
gravity, there is a holographic relation
between FRW brane universe which is reduction from AdS Schwarzschild 
background and boundary CFT which lives on the brane \cite{SV},\cite{SO1}.  
We assume that there are some holographic relations 
between FRW universe which is reduction from dS Schwarzschild
background and boundary CFT. 
To investigate it, we rewrite dS Schwarzschild metric 
(\ref{SAdS}) in the form of FRW metric by using a new time 
parameter $\tau$ following the method of the work \cite{SV}.  
And the parameter $t$ and $r$ in (\ref{SAdS}) are the function 
of $\tau$, namely $r=r(\tau), t=t(\tau)$.  For the purpose of 
getting the 4-dimensional FRW metric, we impose the following 
condition,
\bea
\label{cd1}
-e^{2\rho}\left( {\partial t \over \partial \tau} \right)^2
+e^{-2\rho}\left( {\partial r \over \partial \tau} \right)^2
= -1 \ .
\eea
Thus we obtain FRW metric:
\bea
\label{met1}
ds^{2}_{4}=g_{ij} dx^i dx^j = -d\tau ^2 +r^2 d \Omega_{3}^2 \ .
\eea
The extrinsic curvature, ${\cal K}_{ij}$, of the brane
can be calculated and expressed in term of the function
$r(\tau)$ and $t(\tau)$.  Thus one rewrites
the equations of motion (\ref{eom}) as
\bea
\label{cd2}
{dt \over d\tau}= -{\kappa r  \over 2}e^{-2\rho} \ .
\eea
Using (\ref{cd1}) and (\ref{cd2}), we can derive 
FRW equation for a radiation dominated universe,
Hubble parameter $H$ which is defined by $H={1\over r}{dr \over d\tau}$
is given by 
\bea
\label{HH}
H^2= {1 \over l^2} - {1 \over r^2} + {\mu \over r^4} + {\kappa^2 \over 4}\ .
\eea
Following AdS Schwarzschild gravity case \cite{SV}, 
we choose $\kappa ={2/l}$ from now on \footnote{
From the point of view of brane-world physics \cite{RS}, 
the tension of brane should be determined without ambiguity.
In fact, we  can calculate it to cancel the leading divergence
of bulk AdS Schwarzschild \cite{SO1}.}.  
This equation can be rewritten by
using 4-dimensional energy $E_4$ and volume $V$ 
in the form of the standard FRW equation with
the positive cosmological constant $\Lambda$:
\bea
\label{F1}
H^2 &=& - {1 \over r^2} + {8\pi G_4 \over 3}
{E_4 \over V}+{\Lambda \over 3} \ ,\nn
E_4 &=&{ 3 \mu V  \over 8\pi G_4 r^4},\quad
\Lambda ={ 6 \over l^2 }\ .
\eea
Here $G_{4}$ is the 4-dimensional gravitational
coupling, which is defined by
\bea
\label{gg}
G_{4}={2 G_{5} \over l}\ .
\eea
$E_4$ can be regarded as 4-dimensional 
energy on the brane in dS Schwarzschild 
background which is identical with AdS 
Schwarzschild case.  The cosmological constant $\Lambda$ 
does not appear in AdS Schwarzschild background 
because we can choose brane tension $\kappa$ to cancel
the cosmological constant of AdS Schwarzschild background.
But it is impossible for dS Schwarzschild case 
because if we choose brane tension
to cancel the cosmological constant of dS Schwarzschild background, 
the brane tension should be imaginary. 

By differentiating eq.(\ref{F1}) with respect to $\tau $, 
we obtain the second FRW equation:
\bea
\label{2FR1}
\dot H &=& - 4\pi G_4 \left({E_4 \over V} 
+ p\right) + {1 \over  r^2}\ ,\nn
p &=& {\mu \over 8 \pi G_4 r^4 }\ .
\eea
Here $p$ is 4-dimensional pressure of the 
matter on the boundary.
From eqs.(\ref{F1}) and (\ref{2FR1}), we find that
the energy-momentum tensor is traceless:
\be
\label{trace2}
{T^{{\rm matter}\ \mu}}_\mu=-{E_4 \over V} + 3p 
= 0\ .
\ee
Therefore the matter on the brane can be regarded as
the radiation, which is consistent with the work \cite{CI}
whose calculation has been done in some asymptotically 
dS space in a sense of dS/CFT correspondence.
This result means the field theory on the brane should be 
CFT as in case of AdS Schwarzschild background \cite{SV}.  

Next, we consider space-like brane 
in 5-dimensional dS Schwarzschild background. 
Similarly, we impose the following condition to obtain 
space-like brane metric instead of eq.(\ref{cd1}):
\bea
\label{cd3}
-e^{2\rho}\left( {\partial t \over \partial \tau} \right)^2
+e^{-2\rho}\left( {\partial r \over \partial \tau} \right)^2
= 1 \ .
\eea
Thus we get following FRW-like metric:
\bea
ds^{2}_{4}=g_{ij} dx^i dx^j = d\tau ^2 +r^2 d \Omega_{3}^2 \ .
\eea
Note that this metric is also derived 
by Wick-rotation $\tau \to i\tau$ in eq.(\ref{met1}).
We again calculate the equations of motion
and the extrinsic curvature of space-like brane 
instead of (\ref{eom}) and (\ref{cd2}).  
These equations lead FRW like equation as follows:
\footnote{In the works \cite{DF},\cite{DF1}, the similar equations 
to eqs.(\ref{HH}),(\ref{H3}) were obtained in terms 
of Ricci scalar of the induced metric of the brane.} 
\bea
\label{H3}
H^2= -{1 \over l^2} +{1 \over r^2} - {\mu \over r^4} + 
{\kappa^2 \over 4} \ .
\eea
To cancel the cosmological constant, we take $\kappa ={2/l}$ 
in the same way of AdS Schwarzschild gravity \cite{SV}.  
We assume this equation can be rewritten by
using 4-dimensional energy $E_4$ and volume $V$ 
by the analogous form of the standard FRW equations:
\bea
\label{F3}
H^2 &=& {1 \over r^2} - {8\pi G_4 \over 3}
{E_4 \over V}\ ,\quad E_4 ={ 3 \mu V  \over 8\pi G_4 r^4} \ . \\
\label{3FR1}
\dot H &=&  4\pi G_4 \left({E_4 \over V} 
+ p\right) - {1 \over  r^2}\ ,\quad p={\mu \over 8 \pi G_4 r^4 }\ .
\eea
The reason why the sign of FRW equations
is different from the standard FRW equations (\ref{F1}) 
results from the condition (\ref{cd3}), namely 
$\tau \to i\tau$ in eq.(\ref{met1}).
From eqs.(\ref{F3}) and (\ref{3FR1}), 
we find the energy-momentum tensor is traceless again.  

We stress again that {\it we can take cosmological constant to zero 
for FRW-like equation in space-like brane in
dS Schwarzschild background as the same way in the AdS/CFT 
correspondence.}  This will imply
that the dS/CFT correspondence can be valid for space-like
brane in dS Schwarzschild background.

\section{The Cardy-Verlinde formula for the dS/CFT
correspondence}

E. Verlinde showed that the $d$-dimensional 
FRW equation can be regarded as an analogue of the
Cardy formula of 2-dimensional CFT \cite{EV}.  

\be
\label{CV1}
S_4 =2\pi \sqrt{
{c \over 6}\left(L_0 - {c \over 24}\right)}\ .
\ee
For time-like brane of 5-dimensional dS Schwarzschild 
background, identifying
\bea
\label{CV2}
{2\pi \over 3}\left( E_4 r + {\Lambda V r \over 8\pi G_{4} } 
\right) &\Rightarrow& 2\pi L_0 \ , \nn
{ V \over 8\pi G_4 r} &\Rightarrow& {c \over 24} \ ,\nn 
{ HV \over 2 G_4} &\Rightarrow&  S_4 \ ,
\eea
FRW equation (\ref{F1}) has the form (\ref{CV1}). 
The effect of the cosmological constant 
appears in Cardy formula.  We included contribution
of the cosmological constant in $L_0$ 
because it shifts the vacuum energy.  
This means the cosmological entropy bound \cite{EV} should be changed.  
The Bekenstein bound \cite{EV} in 4-dimensions is 
\bea
S \le S_{B}, \quad  S_{B} \equiv {2\pi \over 3}Er\ .
\eea
Using eq.(\ref{CV2}), the Bekenstein 
entropy bound should be changed as follows:
\bea
S \le S_{B}, \quad  S_{B} \equiv {2\pi \over 3}\left( Er 
+ {\Lambda V r \over 8\pi G_{4} } \right) \ .
\eea
Then we find out that the effect of the cosmological
constant appears in the change of the Bekenstein 
entropy bound. 

For the case of space-like brane, 
identifying
\bea
\label{CV3}
{2\pi \over 3}\ E_4 r  &\Rightarrow& 2\pi L_0 \ , \nn
{ V \over 8\pi G_4 r} &\Rightarrow& {c \over 24} \ , \nn 
-i{ HV \over 2 G_4} &\Rightarrow&  S_4 ,
\eea
Here $H$ changes as $H \to -iH$ since $H$ is defined by
$H={1 \over r}{dr \over d\tau}$ and ${d \over d\tau}$ change as
$-i{d \over d\tau}$ by the Wick-rotation, $\tau\to i\tau$.
The third correspondence in eq.(\ref{CV3}) is identical 
with space-like brane in AdS Schwarzschild background \cite{SO2} exactly.  

The moment when time-like brane crosses 
the black hole horizon\footnote{The time-like brane can only cross
the black hole horizon} $r=r_{H}$ which is derived 
from $e^{2\rho(r_{H})}=0$, the Hubble parameter 
in (\ref{HH}) becomes as $H = \pm 1/l$.  
Here the plus sign corresponds to the expanding brane
universe and the minus one to the contracting universe.  
We choose the expanding case.
Note that the relation $H = \pm 1/l$ is the same form as the case of 
AdS Schwarzschild black hole \cite{SV},\cite{SO1}.  
Using eqs.(\ref{gg}),(\ref{CV2}), we obtain 4-dimensional 
entropy $S_4$ as follows:
\bea
\label{ent}
S_4 ={V \over 2 l G_4}={ V \over 4 G_5}\ .
\eea
This entropy is nothing but the Bekenstein-Hawking 
entropy of 5-dimensional dS black hole similar 
to AdS/CFT correspondence \cite{SV},\cite{SO1}.  

We now understand that discovered relation between FRW
equations and entropy formulas \cite{EV} can be also
applied to dS Schwarzschild background.  If we take
time-like brane which is the boundary
of dS Schwarzschild background, the 
cosmological constant of the brane appears in FRW equations. 
Therefore the effect of cosmological constant 
contributes to raising the the Bekenstein 
entropy bound.  But if we take space-like brane in
dS Schwarzschild background, we obtain the approximately 
same result of AdS Schwarzschild black hole \cite{SV},\cite{SO1}.  
The difference between space-like brane in 
dS Schwarzschild background
and time-like brane in AdS Schwarzschild background is
the sign of FRW equations and third correspondence in 
eq.(\ref{CV3}). When the time-like brane crosses the black hole horizon of 
dS Schwarzschild black hole, the entropy formula of the 
CFT exactly agrees with the black hole entropy of 5-dimensional 
dS background as it happens in 
the AdS/CFT correspondence.  This implies that 
the holographic principle holds true for 
dS Schwarzschild background.

\Acknow
{I would like to thank S. Nojiri, S.D. Odintsov 
for useful discussions. I wish to thank the conference
organizers for their hospitality.  This work is supported in part by 
the Japan Society for the Promotion of Science under the 
Postdoctoral Research Program.
}


\small
\begin{thebibliography}{99}
\bibitem{SOG} S.~Ogushi,\ {\it Mod. Phys. Lett.\/} {\bf A17} (2002) 51, 
hep-th/0111008.  
\bibitem{HL} C.M.~Hull,\ JHEP {\bf 9807}(1998) 021, hep-th/9806146.
\bibitem{AS} A.~Strominger,\ JHEP {\bf 0110}(2001) 034, hep-th/0106113. 
\bibitem{EW} E.~Witten,\ hep-th/0106109,
\bibitem{SO2} S. Nojiri, S.D. Odintsov and S. Ogushi, 
hep-th/0205187 to appear in {\it Int. J. Mod. Phys.}  
Further references are contained therein. 
\bibitem{CI} R.-G.~Cai, Y.S.~Myung and Y.-Z. Zhang, 
{\it Phys. Rev.\/} {\bf D65}, 084019 (2002), hep-th/0110234. 
\bibitem{EV} E.~Verlinde, hep-th/0008140.
\bibitem{CD} J.L.~Cardy, {\it Nucl. Phys.\/} {\bf B270} (1986) 967. 
\bibitem{RS} L.~Randall and R.~Sundrum, {\it Phys. Rev. Lett.\/} 
{\bf 83} (1999) 3370, hep-th/9905221; {\it Phys. Rev. Lett.\/} 
{\bf 83} (1999) 4690, hep-th/9906064. 
\bibitem{SV} I. Savonije and E. Verlinde, 
{\it Phys. Lett.\/} {\bf B 507} (2001) 305, hep-th/0102042.
\bibitem{SO1} S. Nojiri, S.D. Odintsov and S. Ogushi, 
{\it Int. J. Mod. Phys.\/} {\bf A16}(2001) 5085, hep-th/0105117.
\bibitem{SD} S.~Nojiri, S.D.~Odintsov,\ JHEP {\bf 0112}(2001) 033,
hep-th/0107134. 
\bibitem{DF} P.~Binetruy, C.~Deffayet and D.~Langlois, 
{\it Nucl. Phys.\/} {\bf B565}(2000) 269, hep-th/9905012.
\bibitem{DF1} P. Binetruy, C. Deffayet, U. Ellwanger and D. Langlois,
{\it Phys. Lett.\/}{\bf B477}(2000) 285, hep-th/9910219, 
Further references are contained therein.
\end{thebibliography}
\end{document}